\documentclass{aa}  
\usepackage{graphicx}
\usepackage{txfonts}
\begin{document}
   \title{First detection of Zeeman absorption lines in the polar VV Pup}

   \subtitle{Observations of  low activity  states\thanks{Based on
   data collected on the ESO VLT within the program 272.D-5044(A).}}

   \author{E. Mason
          \inst{1}
          \and
          D. Wickramasinghe \inst{2}
	  \and 
	  S. B. Howell \inst{3}
	  \and 
	  P. Szkody \inst{4}
%\fnmsep\thanks{}
          }

   \offprints{E. Mason}

   \institute{European Southern Observatory (ESO)
             Alonso de Cordova 3107, Vitacura, Santiago, CL\\
              \email{emason@eso.org}
         \and
             Australian National University, Australia\\
             \email{dayal@maths.anu.edu.au}
         \and NOAO, 950 N. Cherry Ave., Tucson, AZ USA \\
             \email{howell@noao.edu}
	 \and Dept. of Astronomy, University of Washington, Seattle, WA USA\\
	     \email{szkody@astro.washington.edu}
             }

   \date{Received 3 April 2006 / Accepted  5 March 2007 }

% \abstract{}{}{}{}{} 
% 5 {} token are mandatory
 
  \abstract
    {}
   {We  investigated the low state
   of the polar VV~Pup by collecting high S/N time series spectra.}  
  %  methods heading (mandatory) 
   {We monitored VV~Pup with VLT+FORS1 and
   analyzed  the evolution  of its  spectroscopic features  across two
   orbits.}   
  %  results   heading  (mandatory)  
   {We report the first  detection  of
   photospheric  Zeeman  lines  in  VV  Puppis.   We  argue  that  the
   photospheric  field structure is  inconsistent with  the assumption
   that the accretion shocks are located close to the foot points of a
   closed field line  in a dipolar field distribution.  A more complex
   field structure  and coupling process  is implied making  VV Puppis
   similar  to  other  well  studied  AM Herculis  type  systems.}   
  %  conclusions heading (optional), leave it empty if necessary 
   {}


   \maketitle
%
%________________________________________________________________

\section{Introduction}

VV~Pup  is a  well studied  polar  (see Warner  1995 for  a review  on
polars).  It  was identified as a  polar by Tapia  (1977) who observed
strong and periodic linear  and circular polarization.  VV~Pup was the
first   polar   where  cyclotron   emission   lines  were   identified
(Visvanathan and  Wickramasinghe 1979, Wickramasinghe  and Visvanathan
1980; but see also  Wickramasinghe and Meggitt 1982).  Subsequently, it
was  realized (Wickramasinghe et  al. 1989)  that during  high states,
both  magnetic  poles are  typically  active  and producing  cyclotron
emission.

VV~Pup  has been  observed  at  many different  epochs  and states  of
activity.  Its visual magnitudes,  depending on the state of activity,
vary within  the range  14.5 and 18  (now observed  to go as  faint as
19.5,  see below)  with  values $>$16  mag  all considered  to be  low
states. Though it has already been observed in low states by Thackeray
et al.  (1950), Smak (1971), Liebert et al.  (1978) and Bailey (1978),
this is the first time that  VV~Pup low state is analyzed through time
resolved,  high  S/N, spectroscopy  and,  in  particular, that  Zeeman
absorption lines from the white dwarf are clearly detected.

In this paper  we present the data (Section~2)  and qualitative models
of  the  spectra thus  to  asses the  nature  of  the observed  Zeeman
absorptions (Section~3).   A discussion and a summary  are outlined in
Section~4.

%__________________________________________________________________

\section{Observations}

   \begin{table*}[Hb]
      \caption[]{Log of the observations.}
         \label{log}
     $$ 
         \begin{tabular}{lccccccc}
            \hline
            \noalign{\smallskip}
            Telescope+Inst. & UT date & UT time & shutter time  &  instrument set up & slit width & sky transparency & seeing \\
            \noalign{\smallskip}
            \hline
            \noalign{\smallskip}
	    UT1+FORS1 & 16 Mar 2004 & 23:48-02:23 & 600s$\times$12 & grism 
600V+GG375 & 1'' & CLR & $\leq$1'' \\
	    UT1+FORS1 & 22 Mar 2004 & 01:36-04:11 & 600s$\times$12 & grism 
600V+GG375 & 1'' & CLR & 0.8''-0.9'' \\
            \noalign{\smallskip}
            \hline
         \end{tabular}
     $$ 
%\begin{list}{}{}
%\item[$^{\mathrm{a}}$] This is footnote a
%\end{list}
   \end{table*}

   \begin{figure*}[Ht]
   \centering                            
   \includegraphics[width=8.5cm]{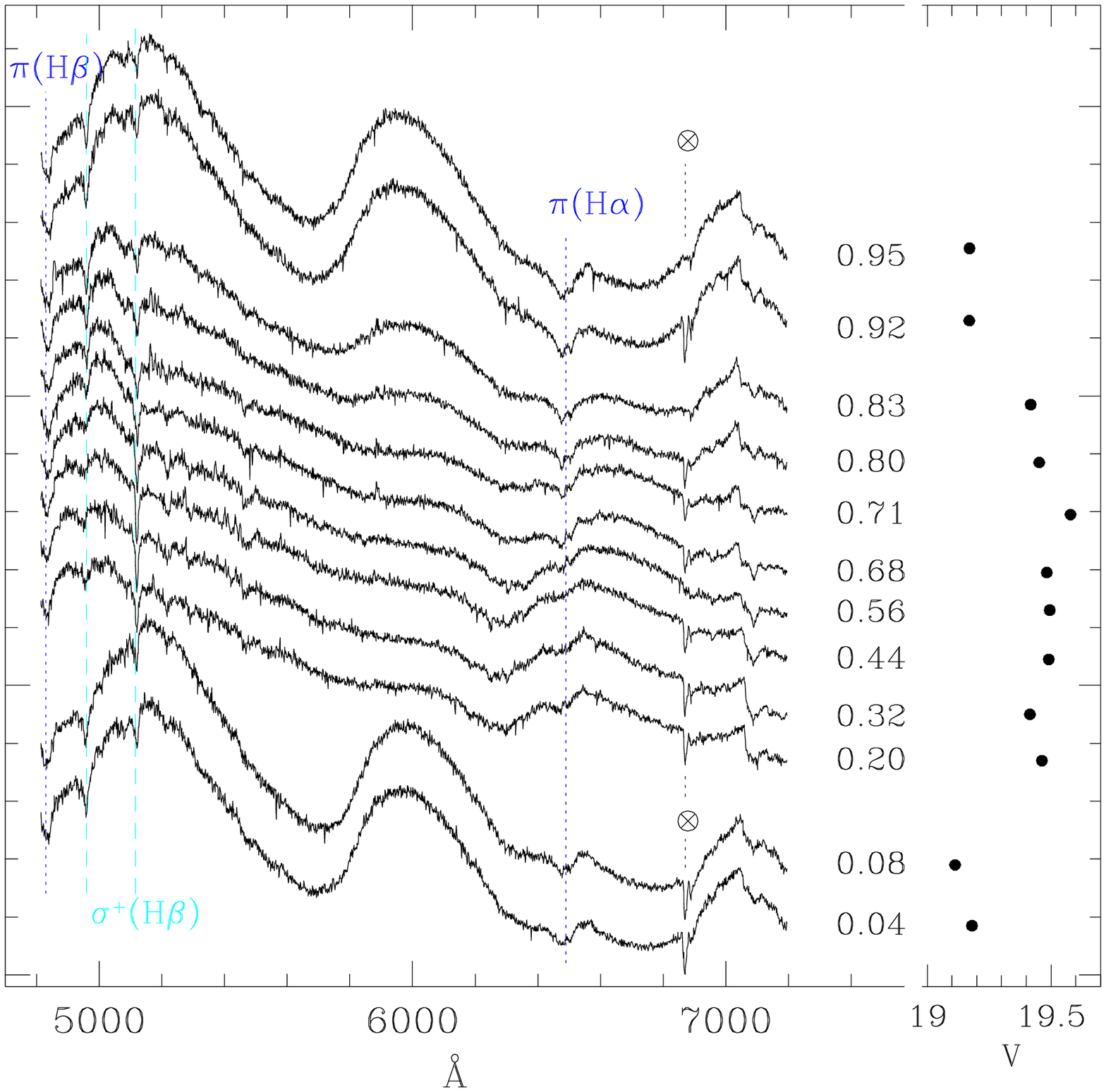} 
   \includegraphics[width=8.5cm]{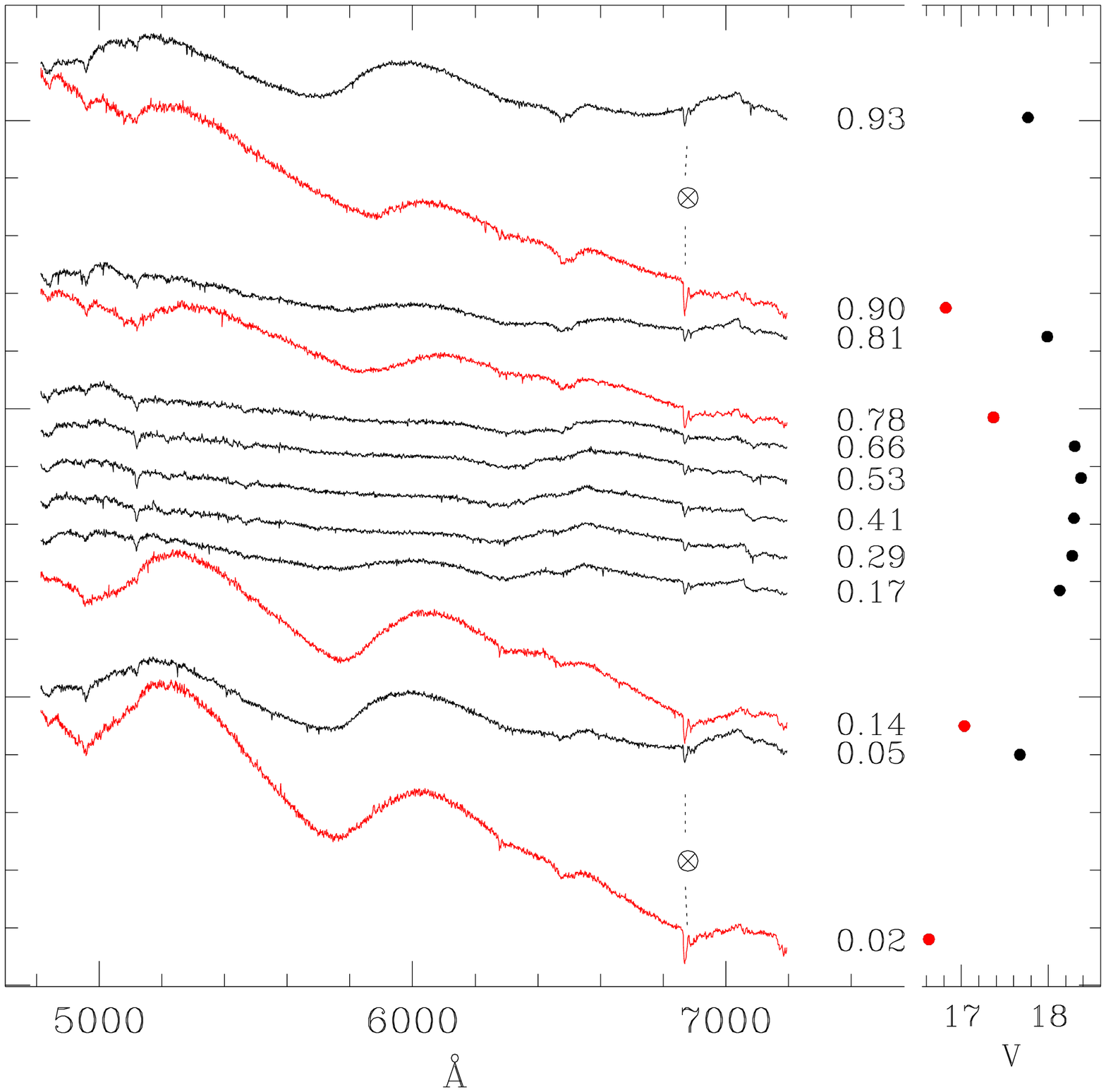}
   \caption{The two  sets of VLT  phase resolved spectra.  Left panel:
   data taken  on 16 March 2004.  Right panel: data taken  on 22 March
   2004.Here, we plot, in red color, the spectra  belonging to a second
   orbital cycle  during which VV~Pup was  considerably brighter. Note
   that the emission  lines have been removed in  order to better show
   the  cyclotron harmonics.   Each spectrum  has been  shifted  by an
   arbitrary constant with respect to the others. We plot on the right
   side of each panel the corresponding light curve.  The V magnitudes
   have  been computed by  convolving each  spectrum with  a V-Johnson
   filter. We calibrated the instrumental magnitude using the standard
   stars LTT~4816  (March 16th) and  EG~274 (March 22nd).   Neither the
   standard stars nor the VV~Pup  spectra have been corrected for slit
   losses. In blue and cyan color we mark, respectively, the $\pi$ and
   $\sigma^+$ Zeeman absorption, from H$\alpha$ and H$\beta$. See text
   for more details.}
              \label{vltSet}
    \end{figure*}

Time  resolved spectroscopy  across  an orbital  period was  performed
 twice, on  the 16th and the  22nd of March 2004  using VLT+FORS1. The
 observations were secured  though Director's Discretionary Time (DDT)
 shortly after  VV~Pup was observed to  be in a low  state.  The spectra
 were taken  in Long Slit Spectroscopy  (LSS) mode using  slit 1'' and
 grism 600V together with the  order sorting filter GG375. The covered
 wavelength range  is 4800-7300 \AA \ with  linear spectral dispersion
 of $\sim$1.18 \AA/pix. Each data set consists of 12 spectra of 10 min
 exposure each.  Table~\ref{log} reports the log of the observations.

Data reduction (bias subtraction, flat correction, wavelength and flux
calibration)  was performed according  to standard  procedures through
IRAF  routines.   All  the  spectra  were  subsequently  phased  using
Walker's  (1965)  photometric  ephemeris.   Note  that  Walker's
ephemeris was based  on the time of maximum  light observed in optical
broad  band photometry.  His  phase  zero agrees  with  that for  the
secondary  star inferior conjunction  to within  $\pm$0.02 of  a phase
(Howell et al.  2006a).  
Given the rough agreement of Walker's ephemeris with the recent radial
velocity determinations,  we have used it herein  for consistency with
previous work on VV~Pup.

\section{Data analysis and modeling}
\label{theo}

   \begin{figure*}[Ht]
   \centering
   \includegraphics[width=8.5cm]{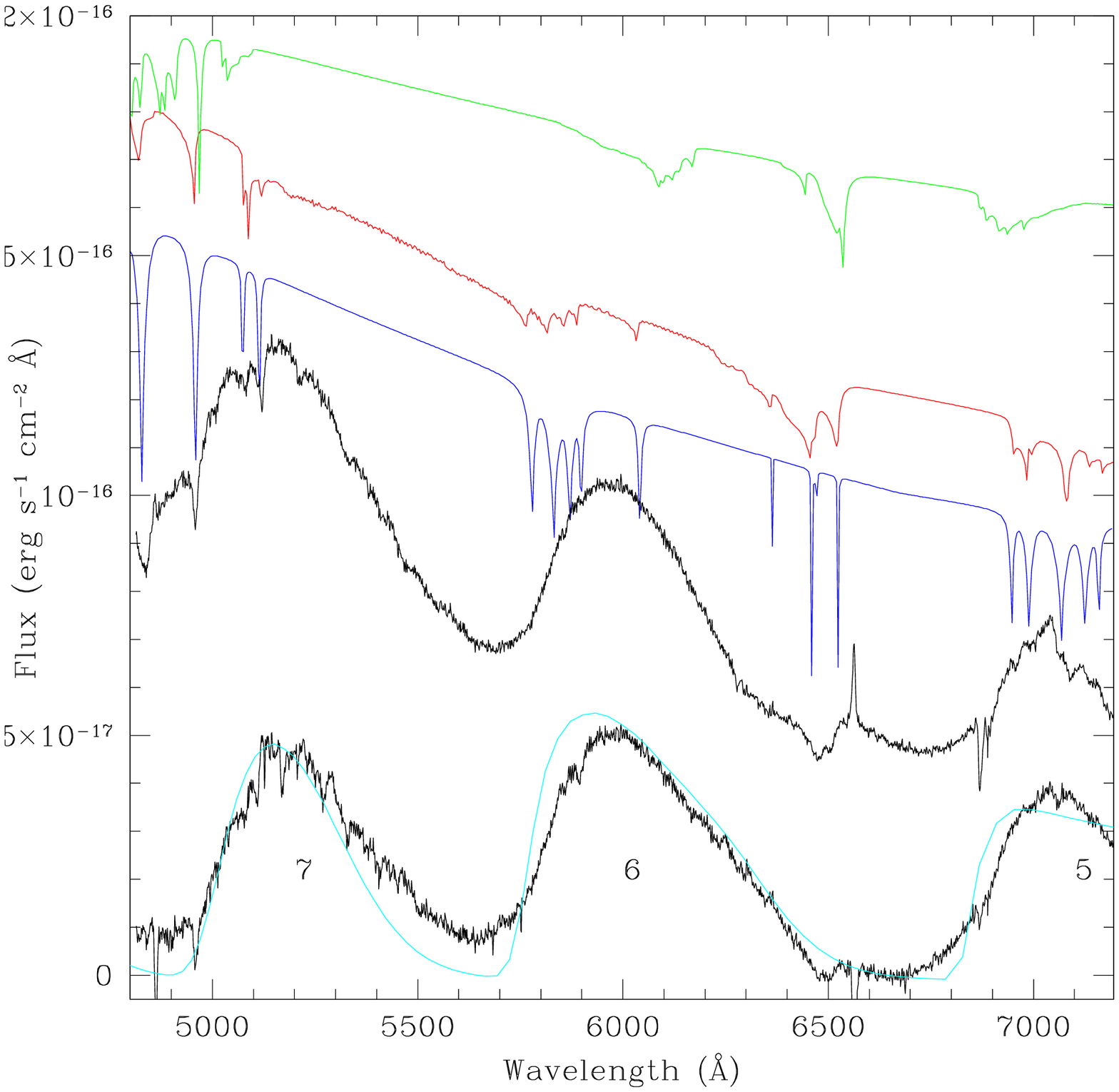}
   \includegraphics[width=8.5cm]{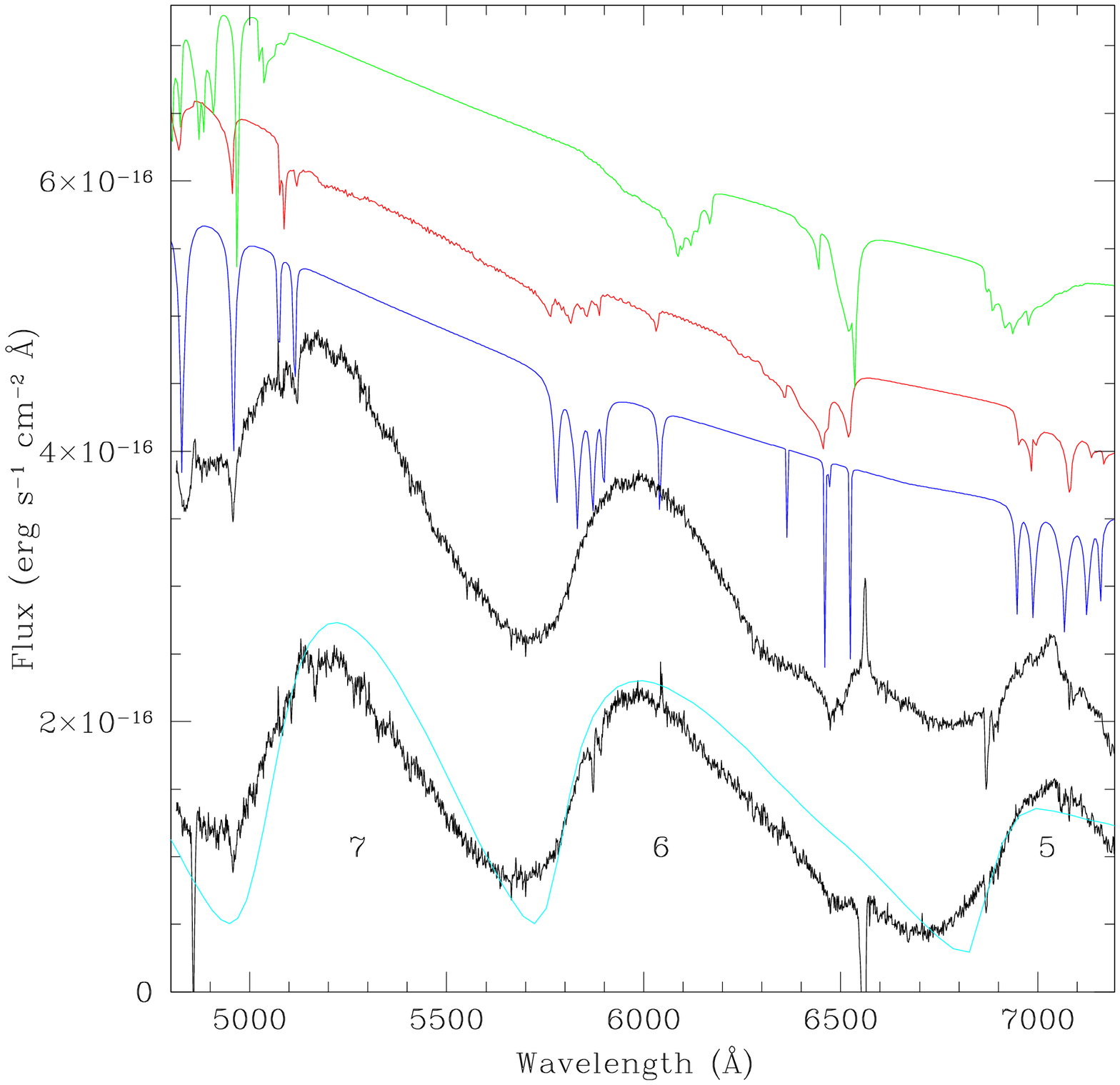}
      \caption{Cyclotron  (cyan line),  halo Zeeman  ($B_{h1}= 32$~MG,
blue  line), and  the  two photospheric  models:  $B_d=40 MG,  d=0.11,
\theta=105^\circ$  (off center  dipole, green)  and  $B_d=65MG, d=0.0,
\theta=55^\circ$  (centered dipole,  red)  discussed in  the text  and
compared  with spectra (black  lines) from  the data  set taken  on 16
March 2004  (left panel) and on  22 March 2004 (right  panel). In each
panel, the observed spectrum on  the top is an average VV~Pup's bright
phase spectrum;  while the  spectrum on the  bottom is  the difference
between the ``average bright  phase spectrum'' and the ``average faint
phase''  one.  For  the  data set  of  March 16  ``the average  bright
spectrum'' consists of the average of the four spectra between orbital
phase 0.92  and 0.08; while the  ``faint spectrum'' is  the average of
the two spectra  at orbital phase 0.44 and 0.56.  For  the data set of
March 22  the ``average  bright spectrum'' is  the average of  the two
spectra at orbital phase 0.05  and 0.93; while, the ``faint spectrum''
results  from the average  of the  spectra at  orbital phase  0.53 and
0.66.  The numbers below the ``difference spectra'' mark the cyclotron
harmonics. Flux  units correspond to the observed  spectra. The models
are in arbitrarily scaled units.  }
         \label{Model16}
   \end{figure*}  

\subsection{March 2004 cyclotron spectra}
\label{2pole}

We modeled the cyclotron emission we observe in the two time series of
FORS1 spectra.   Though cyclotron modeling of VV~Pup  has already been
performed in  the past deriving the  magnetic field at  the two poles,
here we  repeat the analysis  both to search for  possible differences
and to use our newly confirmed results to constraint the dipole models
when interpreting the Zeeman spectra.

We plot in Figure~\ref{vltSet} the two series of spectra after phasing
and after having  removed the  emission lines to  enhance the  visibility of
cyclotron   emissions.   The  second   data   set  probably
corresponds  to  a  higher  mass  transfer rate,  being  about  1  mag
brighter. Clear evidence of a  different mass transfer rate is present
in the emission  lines (intensity, species and profile)  which will be
analyzed in a separate paper. However, according to the AAVSO records,
VV~Pup  was never  brighter than  V=16 mag,  thus both  our  data sets
sample a low state.

In  March 2004,  VV Pup  was clearly  accreting through  both magnetic
  poles  as it is  evident from  the two  distinct sets  of cyclotron
  emission/humps  which  are  visible  at different  orbital  phases.
  During  the bright  orbital  phase ($\sim$0.8-0.2)  the spectra  are
  dominated by  cyclotron emission peaking at  5170~\AA, 6000~\AA, and
  7050~\AA.   These cyclotron  harmonics  are attributed  to the  main
  $\sim 32$~MG pole.  During  the faint orbital phase ($\sim$0.2-0.8),
  these harmonics  disappear 
and  give  way to
  another series of cyclotron peaks  at 5000~\AA \ and 6600~\AA.  This
  second set  forms in the proximity  of the stronger  ($\sim 56$~ MG)
  magnetic pole.  Based  on the assumption that the  secondary pole is
  visible at all  phases (see, for example, Piirola  et al.  1990), we
  have  subtracted the  faint  phase spectrum  from  the bright  phase
  spectrum to extract what we expect for the main pole only.

The  cyclotron  models assume  the  point  source  model described  in
Wickramasinghe and Meggitt (1985). Each model is defined by a set of 4
parameters corresponding to the polar magnetic field, $B_i$ (where $i$
is  either 1 or  2), the  electron temperature,  T, the  optical depth
parameter, $\Lambda$, and the tilting angle, $\delta$, between the field
direction and the  line of sight.  The parameters  that best reproduce
the main pole spectrum are $B_1=31.5$~MG, T=5 Kev, $\log \Lambda=5.5$,
$\delta=85^o$  for the  data  taken on  the  16th of  March and  $B_1=
31.5$~MG, T=7  Kev, $\log \Lambda=5.3$, $\delta=85^o$  for those taken
on the 22nd of March  (see Figure~\ref{Model16}, cyan lines).  We note
that the cyclotron spectrum  shows little flux between harmonics $n=5$
and  $n=6$  during  March  16,  consistent  with  the  lower  electron
temperature of  this model.  We emphasize  that the values  of $T$ and
$\Lambda$ derived  from point source models are  indicative of average
values over an extended shock  region and cannot be directly converted
to fluxes without consideration of geometrical effects.

   \begin{figure}[h]
   \centering
   \includegraphics[width=8.5cm]{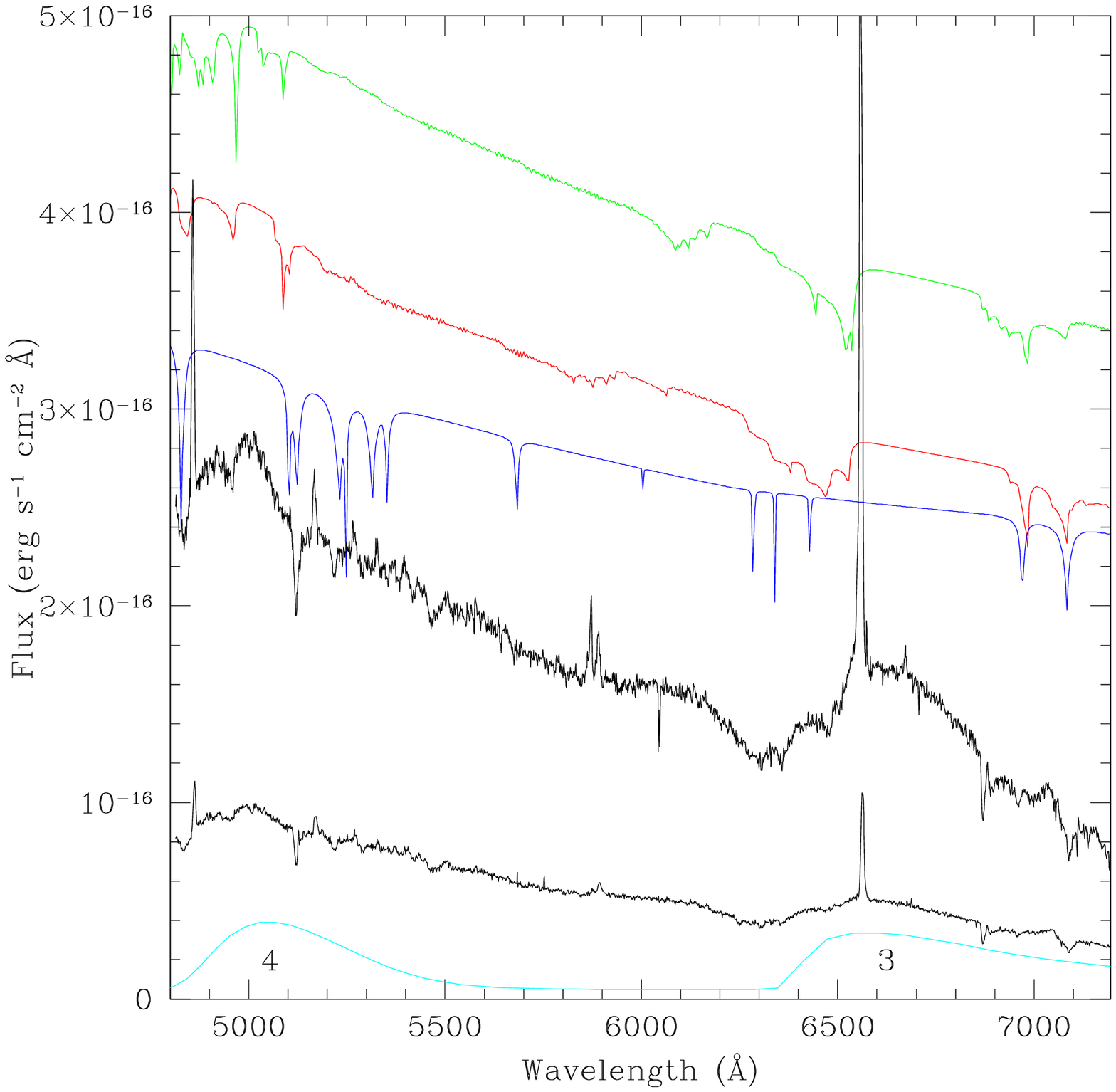}
      \caption{The  faint   phase  spectra  (black   lines)  taken  at
      VLT+FORS1  on the  16th (bottom)  and 22nd  (top) of  March 2004
      together with  the cyclotron model (cyan line).   Also shown are
      possible  contributions from  a halo  with  $B_{h2}=56$~MG (blue
      line) due to  the second magnetic pole and  from the photosphere
      (green and red lines).  The photospheric models have $B_d=40 MG,
      d=0.11,  \theta=45^\circ$  (green  line) and  $B_d=60MG,  d=0.0,
      \theta=35^\circ$ (red  line).  See  text for more  details.  The
      faint spectrum of March 16 is  the average of the two spectra at
      orbital phase 0.44  and 0.56. The faint spectrum  of March 22 is
      the average of  the two spectra at orbital  phase 0.53 and 0.66.
      Flux units correspond to the observed spectra, while, the models
      are in arbitrarily scaled units.}
         \label{vltFaint}
   \end{figure}

The photosphere is expected to make important contributions during the
faint phase, thus it is harder to isolate  the properties  of the
cyclotron component from the faint phase emission. The position of the
harmonics, however,  allows us to  estimate the magnetic field  at the
second pole  with good accuracy,  even though the viewing  geometry of
this region is  not well established because it  is not eclipsed. The
expected positions and relative strengths of the cyclotron lines for a
model    with   $B_2=54.6$~MG,    T=5    Kev,   $\log    \Lambda=3.0$,
$\delta=75^\circ$ are shown  in Figure~\ref{vltFaint}. The lower value
of this field  compared to what was deduced  in the discovery spectrum
of the second  pole (Wickramasinghe et al. 1989)  is significant and
must indicate a change in the accretion geometry.

   \begin{figure*}
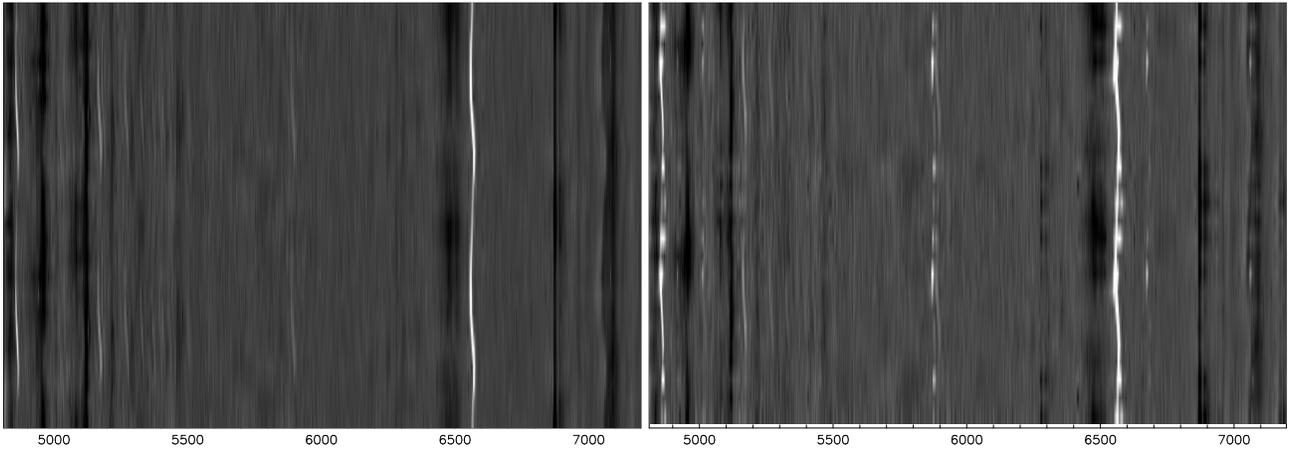

   \centering         \includegraphics[angle=-90,width=8.5cm]{5354fig4a.ps}
   \includegraphics[angle=-90,width=8.5cm]{5354fig4b.ps}
   \caption{Trailed  spectrograms of the  two VLT  data sets.   On the
   right panel the data set taken  on 16 March 2004; on the left panel
   the set  taken on  22 March 2004.   Spectra are phase  ordered from
   bottom to top and the  orbital cycle has been repeated for clarity.
   However, the phase spacing between two consecutive spectra is not a
   constant value.  Note that the spectra  showing stronger H$\alpha$
   and     H$\beta$    emission     as    well     as     the    blend
   HeI(11)$\lambda$5876+NaI$\lambda$5890,5896, correspond  to a second
   orbital  cycle, of  March 22  observations, which  was considerably
   brighter than the first orbital cycle. }
              \label{lineTrail}
    \end{figure*}

\subsection{Zeeman spectra}

Narrow  absorption   features  which  can  be   identified  as  Zeeman
components of  $H\alpha$ and $H\beta$  are clearly seen  in both
sets of  the March data (Figure~\ref{vltSet}).  Such  features are seen
in many  other AM  Herculis type systems  and are  usually interpreted
either as being  of photospheric origin or as  evidence for absorption
in localized regions associated with an accretion shock (an accretion
halo).

It is clear from  Figure~\ref{vltSet} that the absorption features are
seen  at all phases  at nearly  the same  wavelength. We  identify, in
particular, the  features shortward of  $5200$ \AA \  with $H\beta$
Zeeman components  and the structured  feature near $6490$ \AA  \ with
the $\pi$ component of  $H\alpha$.  The trailed spectra reported in
Figure~\ref{lineTrail}  shows  that   absorption  line  profiles  vary
considerably across  the orbit. The  line profiles change  from narrow
and sharp during the bright phase to shallow and weak during the faint
phase and  this prevents  the detection of  any clear orbital/sinusoidal
motion.  Line  shifts due  to changes in  the field structure  are not
expected to  be correlated with radial velocity.   Their visibility at
all phases  would suggest that  the dominant contribution is  from the
photosphere.  However, the  $\sim 56$ ~MG pole is  seen at all phases,
and may contribute  narrow halo lines, and the $\sim  32$ ~MG pole may
contribute narrow halo lines during the bright phase.

In  order to  assess whether  those absorption  lines originate in the
photosphere  or the  halo, we  proceed  by comparing  our spectra  with
Zeeman absorption as predicted by halo and photosphere models.

There are  no detailed theoretical  models of the thermal  and density
structure of  accretion halos.  An  indication of the type  of spectra
that may  be expected from a  halo can be obtained  by considering the
spectra from  photospheric patches  in a nearly  uniform field  on the
stellar surface in the hypothesis that the halo lines are formed in an
effective reversing layer associated with  cool gas in the vicinity of
the shock. We present the  results of such calculations for polar caps
with  $B_{h1}=32$~MG,  $B_{h2}=56$~MG  with  a field  spread  of  $10$
percent in Figure~\ref{Model16} and \ref{vltFaint}, respectively (blue
lines).   Of  course, the  relative  strengths  of  the $H\alpha$  and
$H\beta$ transitions may be quite different from predictions of these
models  (see  Achilleos  and  Wickramasinghe 1989)  depending  on  the
ionization  and   excitation  conditions  in  the   halo.   A  general
observation  that applies  to all  the VLT  spectra including  the two
shown in Figure~\ref{Model16} and \ref{vltFaint} is that the widths of
the  dominant  $\pi$ components  of  $H\alpha$  appear  to be  much
broader  than  predicted  by   the  halo  models.   Also,  the  narrow
H$\alpha$~$\sigma^-$ component  that we expect at $5700$~\AA  \ from a
$56$~MG  halo,   is  not  present   in  the  faint  phase   data  (see
Figure~\ref{vltFaint}).  These two arguments  alone disfavor  the halo
interpretation of the Zeeman lines.

Conversely, the narrowness of the H$\beta$ absorption features (see in
particular the  bright phase  spectra of Figure~\ref{Model16}  and the
absorption lines at 4840, 4960  and 5118~\AA), seems to support a halo
interpretation,  if one  assumes that  the appropriate  ionization and
excitation conditions prevail in  this region. However, it is unlikely
that these  are halo features from  the main $32$~MG  pole because the
features associated with the blend of the five $\sigma^{-}$ components
due to  $H\alpha$ near $5800$  ~\AA \ is  not present in  our data.
Since halo lines are formed {\it against} a cyclotron component, these
lines should be clearly seen in the data.

We next  consider a photospheric interpretation for  the Zeeman lines.
Off centered dipole models are often used for modeling the underlying
field structures in  white dwarfs. We consider such  models as a first
approximation to the field  structure. The models are characterized by
the  dipolar field  strength $B_d$,  the dipole  offset $d$  along the
dipole  axis, the  tilt  of  the dipole  axis  $\delta_{d}$, and  the
orbital inclination $i$ (see e.g. Wickramasinghe and Martin 1979).

We present two  sets of dipole models.  The first set  is based on the
hypothesis  that the  cyclotron field  determinations can  be  used to
constrain  field  structure.   Previous  studies  of  photometric  and
polarimetric variations of VV~Pup have constrained the location of the
cyclotron  emission regions  on the  stellar surface  and  the orbital
inclination  $i$.  If  one  makes the  assumption  that the  accretion
shocks are located close to the magnetic poles at the foot points of a
closed field line,  and the field structure is  dipolar, the different
field  strengths measured from  the cyclotron  lines would  indicate a
dipole of strength $B_d= 40$~MG with an offset $d \sim 0.11$ along the
dipole  axis. Indeed,  such a  model  gives polar  field strengths  of
$57$~MG  and $29$~MG.   This model  requires  a tilt  of the  magnetic
dipole axis of  $\delta_{d}\sim 30^\circ$ to the rotation  axis and an
orbital inclinations  $i\sim 75^\circ$ in order to  be consistent with
the eclipse  and polarization constraints  (Meggitt and Wickramasinghe
1989).  As the dipole rotates, the angle between the line of sight and
the dipole  axis varies between  $\theta=105^\circ$, at the  center of
the bright  phase, and $\theta=45^\circ$,  at the center of  the faint
phase.  The expected faint and  bright phase spectra from such a model
are  shown in Figure~\ref{Model16}  and \ref{vltFaint}  (green lines).
These models are clearly also  not consistent with the faint or bright
phase  observations.  The  central, and  strongest  $H\alpha$ $\pi$
component is shifted redward from the observed position, and the blend
of $H\alpha$  $\sigma^-$ components predicted near $6100$  \AA \ is
not present in the observations.   This is particularly evident at the
viewing angle of $\theta=105^\circ$ when the more uniform field region
containing the weaker pole is in view. The obvious implication is that
the two  key assumptions involved  in these models (namely  the closed
field  line and  dipolar  field approximation),  when taken  together,
yield inconsistent  results.  Previous investigators  have noted phase
shifts  between the  relative  positions of  the  accretion shocks  at
different states of activity and have used cyclotron spectroscopy also
to argue against  the closed field line assumption  (e.g.  Schwope and
Beuermann 1997).
 
A  second alternative approach  consists in  admitting that  the field
structures  cannot be  constrained by  cyclotron  field determinations
(they only provide measurements of  the field in two localized regions
on the white dwarf surface), and the manner in which the mass transfer
stream  couples to  field  lines is  essentially  unknown.  The  field
structure could then be  constrained by analyzing the phase dependence
of the  absorption lines in conjunction  with spectropolarimetric data
(e.g. Beuermann et  al. 2007). However, such studies  are usually best
carried out  using data  obtained when  the system is  in a  truly low
state.

We conclude this section by noting that the data currently at hand can
nevertheless be used  to provide some very general  constraints on the
underlying field  structure. If we restrict our  investigations to the
class  of centered  dipole  models, and  investigate  the spectrum  at
different angles, $\theta$,  to the white dwarf rotation  axis, we can
conclude that there  is not a unique centered  dipole model capable of
explaining all the observations.  However, we can find reasonable fits
to the data  at the center of  the bright phase and the  center of the
faint phase by allowing  different values of $B_d$.  This demonstrates
that a field structure that is  more complex than a centered dipole is
required to model VV~Pup spectra.   These models are compared with the
observations   in    Figures~\ref{Model16}   ($B_d   =    65$~MG   and
$\theta=55^\circ$)     and    \ref{vltFaint}    ($B_d     =    60$~MG,
$\theta=35^\circ$), respectively.

\section{Summary and conclusions}

We have presented two time series  of spectra taken during a low state
of the  polar VV~Pup.  Each time series  cover about 1  orbital period
with  a  phase  resolution  of  0.1 per  spectrum.   The  spectra  are
characterized  by  very few  emission  lines  (signature  of low  mass
transfer rate), cyclotron emission and Zeeman absorption lines.

In order  to identify  the exact nature  of the Zeeman  absorptions we
made  use of  simple  models representative  of  halo or  photospheric
Zeeman  absorptions.   In  the  first  case we  have  shown  that  the
predicted  absorptions are  more than  the observed  ones  and/or have
widths which  are inconsistent with  observations. On the  other side,
also the photospheric dipole model predicts a Zeeman spectrum which is
in serious  disagreement with  the observations.  In  particular, they
predict redder  absorption lines than observed (off  center dipole) or
spectral  changes with  the rotational  phase that  we do  not observe
(centered dipole).  We conclude that the actual VV~Pup field is likely
a  complex non-dipolar  one. We  note that  multi-pole fields  are not
uncommon  among both  isolated magnetic  white dwarfs  and  polars. In
particular, complex field structures have been modeled for polars such
as  EF  Eri,  BL  Hyi  and  MR Ser  (Reincsh  et  al.  2005),  through
spectropolarimetry.   A better understanding  of the  underlying field
distribution must await detailed spectropolarimetric observations when
VV Puppis descends into a truly  low state and the bare photosphere is
revealed.

Our observations provided  clear evidence for accretion onto both
magnetic poles.  By modeling  the cyclotron emissions we find magnetic
fields  of  $31.5$~MG  and  $54.6  $~MG for  the  main  and  secondary
accretion  region, respectively.  We note  that   two pole
accretion during  a very low state  has not been  reported before.  In
the  past VV~Pup  has been  observed to  accrete  through  both the 
magnetic   poles   (e.g.    Visvanathan   and   Wickramasinghe   1981,
Wickramasinghe et al.   1989, Schwope and Beuermann 1997),  as well as
just the main one  (e.g.  Wickramasinghe and Visvanathan 1980, Canalle
and  Opher 1988,  Wickramasinghe et  al.  1984),  during  high states.

Polars  are known to  exhibit different  observational characteristics
depending on the state of activity of the system. The different states
are  usually associated  with evidence  of accretion  onto one  or two
poles, or  a combination of  the two (see Wickramasinghe  and Ferrario
2000, for  a review). However,  in the case  of VV~Pup, it  seems that
there is  no correlation between  the accretion configuration  and the
state/ongoing mass  transfer rate.  This fact  questions the mechanism
controlling the accretion geometry. There could well be other factors,
independent on the mass transfer  rate (e.g.  star-spots and or active
regions  on the secondary  star, see  Howell et  al 2006b),  which are
capable of  triggering the two pole  accretion.  Alternatively, VV~Pup
could  have undergone  a change  in  the relative  orientation of  the
magnetic axis.   The latter possibility has already  been reported for
DQ  Leo (Wickramasinghe  and  Ferrario 2000),  for example.   However,
particularly to prove or disprove the hypothesis of secular variations
of  the  magnetic  axis  orientation,  long  term  spectropolarimetric
observations are needed and encouraged.

\begin{acknowledgements}
The  authors  are grateful  to  the  ESO  'Director General'  for  the
allocation of  the VLT time  allowing the observations. SBH  wishes to
thank the ESO/Santiago Visiting  Scientist Program for the approval of
a scientific visit during which this paper was completed.

\end{acknowledgements}


\begin{thebibliography}{}
\bibitem[ac]{1989} Achilleos, N., Wickramasinghe, D. T., 1989, ApJ, 346, 444

\bibitem[A]{1982} Allen, D. A., Cherepashchuk, A. M., 1982, MNRAS, 201, 521

\bibitem[bailey]{1978} Bailey, J., 1978, MNRAS, 185, {\it Short Communication}, 
73

\bibitem[beu]{2007}  Beuermann, K., Euchner, F., Reincsh, K., Jordan, S., Gaensicke, B.T., 2007, A\&A, accepted (astro-ph/0610804)

\bibitem[CO]{1988} Canalle, J. B. G., Opher, R., 1988, A\&A, 189, 325

\bibitem[CCH]{1982} Cowley, A. P., Crampton, D., Hutchings, J. B., 1982, 259, 
730

\bibitem[DS]{1994} Diaz, M. P., Steiner, J. E., 1994, A\&A, 283, 508

\bibitem[GG]{1978} Giampapa, M., et al., 1978, ApJ, 226, 144

\bibitem[har]{2005} Harrison, T. E., Howell, S. B., Szkody, P., Cordova, F. A., 
2005, ApJ, 632, L123 

\bibitem[h]{2001} Howell, S. B., Gelino, D., and Harrison, 2001, AJ, 121, 482

\bibitem[h]{2002} Howell, S. B., Ciardi, D., Sirk, M., and Schwope, A., 2002, 
AJ, 123, 420

\bibitem[h]{2006}  Howell, S. B.,  Harrison, T.  E., Campbell,  R. K.,
Cordova, F. A., Szkody P., 2006a, AJ, 131, 2216

\bibitem[h]{2006} Howell, S. B., Walter, F., Harrison, T. E., Huber, M. E., 
Becker, R. H., White, R. L., 2006b, ApJ, 652, 709

\bibitem[ks]{2005}  Kafka,  S.,  Honeykutt,  R.  K.,  Howell,  S.  B.,
Harrison, T. E., 2005, AJ, 130, 2852

\bibitem[ks2]{2006} Kafka, S. Honeycutt, R. K, Howell, S. B., 2006, AJ, in press

\bibitem[LiSt]{1979} Liebert, J., Stockman, H. S., Angel, J. R. P., Woolf, N. 
J., Hege, K., Margon, B.,  1978, ApJ, 225, 201

\bibitem[Libe]{2003} Liebert, J., Kirkpatrick, J. D., Cruz, K. L., Reid, I. N., Burgasser, A., Tinney, C. G., Gizis, John E., 2003, AJ, 125, 343

\bibitem[mw]{1989} Meggitt, S. M. A., Wickramasinghe, D. T., 1989, MNRAS, 236, 
31

\bibitem[pandel]{2005} Pandel, D., Cordova, F. A., ApJ, 620, 416

\bibitem[patt]{1984} Patterson, J., Lamb, D. Q., Fabbiano, G., Raymond, J. C., Beuermann, K., Swank, J., White, N. E, 1984, ApJ, 279, 785 

\bibitem[piirola]{1990} Piirola, V., Coyne, G. V., Reiz, A., 1990, A\&A, 235, 245

\bibitem[Reincsh]{2005} Reincsh, K.,Euchner, F., Beuermann, K., Jordan, S.,  
Gansicke, B. T., 2005, ASP Conf. Ser  330, 177

\bibitem[SY]{1980} Schneider, D. P., Young, P., 1980, ApJ, 240, 871

\bibitem[Schwope]{1997} Schwope, A. D., Beuermann, K., 1997, AN, 318, 111

\bibitem[Schp]{2001} Schwope, A. D., Schwarz, R., Sirk, M., Howell, S. B., 2001, A\&A, 375, 419

\bibitem[sh]{1998} Sirk, M. and Howell, S. B., 1998, ApJ, 506, 824 

\bibitem[smak]{1971} Smak, J., 1971, AcA, 21, 467

\bibitem[Sz]{1983} Szkody, P., Bailey, J. A., Hough, J. H., 1983, MNRAS, 203, 
749

\bibitem[1977]{tapia} Tapia, S., 1977, IAUC N. 3054

\bibitem[1950]{thac} Thackeray, A. D., Wesselink, A. J., Oosterhoff, P. Th.,  
1950, BAN, 11, 193

\bibitem[1985]{uch} Uchida, Y. and Sakuri, T., 1985, IAUS, 107, 281

\bibitem[1979]{VW79} Visvanathan, N., Wickramasinghe, D. T., 1979, Nature, 281, 
47

\bibitem[1981]{VW81} Visvanathan, N., Wickramasinghe, D. T., 1981, MNRAS, 196, 
275

\bibitem[1988]{Wh} Wade, R., and Horne, K., 1988, ApJ, 324, 411

\bibitem[1965]{walker} Walker, M. F., 1965, Communication of the Konkoloy 
Observatory, 57, 1

\bibitem[1995]{w95} Warner, B., 1995, {\it Cataclysmic variable stars}, 
Cambridge Univeristy Press

\bibitem[1980]{WV80} Wickramasinghe, D. T., Visvanathan, N., 1980, MNRAS, 191, 
589 

\bibitem[1982]{WM82} Wickramasinghe, D. T., Meggitt, S. M. A.,  1982, MNRAS, 
198, 975

\bibitem[1979]{WM} Wickramasinghe, D. T., Martin, 1979, MNRAS, 188, 165

\bibitem[1985]{WM85} Wickramasinghe, D. T., Meggitt, S. M. A.,  1985, MNRAS, 
214, 605

\bibitem[1989]{W_2pole} Wickramasinghe, D. T., Ferrario, L., Bailey, J., 1989, 
ApJ, 342, L35 
   
\bibitem[WRB]{1984} Wickramasinghe, D. T., Reid, I. N., Bessel, M. S., 1984, 
MNRAS, 210, {\it Short Communication}, 37

\bibitem[1989]{rev} Wickramasinghe, D. T., Ferrario, L., 2000, PASP, 112, 873 
   

\end{thebibliography}
\end{document}